\documentclass[12pt]{aa}
\usepackage[english]{babel}
\usepackage{graphicx}

\onecolumn
\begin{document}

\title{Gusty wind in the system of the infrared source RAFGL\,5081}

\author{V.G.\,Klochkova, E.L.\,Chentsov, V.E.\,Panchuk,  N.S.\,Tavolzhanskaya, and M.V.\,Yushkin }

\institute{Special Astrophysical Observatory RAS, Nizhnij Arkhyz, 369167 Russia}

\date{\today} 

\abstract{For the first time, based on long-term spectral monitoring with high spectral resolution, the optical 
spectrum of the weak central star of the IR source RAFGL\,5081 has been studied. The spectral
type of this star is close to G5$\div$8\,II, and its effective temperature is Teff$\approx$5400\,K. An unusual 
spectral phenomenon was discovered: splitting of the profiles of broad, stationary absorption lines of medium and
low intensity. The heliocentric radial velocities Vr of all components of metal absorption lines, the Na\,I~D lines, 
and the H$\alpha$ line were measured for all the observation epochs. The constancy of the absorption
lines rules out the possibility that the line splitting is due to binarity. The radial velocities of the wind
components in the profiles of the Na\,I~D and H$\alpha$ lines reach $-250$ and $-600$\,km/s, respectively. 
These profiles have narrow components, whose number, depth, and position vary with time. The time variability
and multicomponent structure of the profiles of the Na\,I~D and H$\alpha$ lines indicates inhomogeneity and
instability of the circumstellar envelope of RAFGL\,5081. The presence of components with velocity
Vr(IS)\,=$−65$\,km/s in the Na\,I\,(1) lines provides evidence that RAFGL\,5081 is located behind the Perseus arm, 
i.e, no closer than 2\,kpc. It is noted that RAFGL\,5081 is associated with the reflection nebula GN\,02.44.7. 
\newline
{\it Keywords: stars, evolution, AGB stars, circumstellar envelopes, optical spectroscopy.} }

\authorrunning{\it Klochkova et al.}
\titlerunning{\it Gusty wind in the system of the infrared source RAFGL\,5081}

\maketitle

\section{Introduction}

As its name implies, the infrared (IR) source RAFGL\,5081 (=IRAS 02441+6922) was detected in the earliest balloon 
IR surveys.  Later, Kwok et al.~[1] included it in a list of candidate asymptotic giant branch (AGB)
stars. Evolved AGB stars and stars in the post--AGB stage, observed during the short evolutionary transition 
to a planetary nebula (protoplanetary nebulae, PPN), have low-mass cores with masses of about 0.6\,M$_{\sun}$. 
The degenerate cores are surrounded by extended gas--dust envelopes that formed due to the substantial loss 
of stellar matter in earlier evolutionary stages. The presence of circumstellar gas and dust is manifest 
by the characteristic infrared, radio, and optical spectra of PPN. 

The optical spectra of these low-mass supergiants differ from the spectra of classical massive supergiants 
in the presence of molecular bands superimposed on the  spectrum of a cool supergiant, anomalies in the profiles 
of the H\,I, Na\,I, and He\,I absorption and emission  lines, and the presence of emission lines of some metals 
(for details and references, see~[2]). In addition,  all these spectral features often vary with time. 
Overall, we observe the following features in the optical  spectra of post-AGB  supergiants, which distinguish 
them from the spectra of massive supergiants: complex  profiles of H\,I containing absorption and emission 
components that vary with time, absorption or emission  molecular bands, mainly of carbon-containing molecules, 
envelope components of Na\,I and K\,I resonance  lines, narrow forbidden or permitted emissions  of metals formed 
in the envelope.

To date, there is no information about the central star of RAFGL\,5081. Hrivnak et al.~[3] did not 
find an optical counterpart of RAFGL\,5081 in their study on the identification of OH/IR objects. 
There are neither spectral nor photometric observational data for this object in the visual in the 
SIMBAD database. There are also very few publications about observations of RAFGL\,5081 in the other 
wavelength ranges. Using the results of IR observations on the Canada--France--Hawaii telescope and IRAS data, 
Kwok et al.~[1] classified RAFGL\,5081 as a member of a the group of sources with their maximum IR radiation at
$\lambda < 10$\,$\mu$, but could not say anything definite about the nature of the object. No radiation 
is detected in CO, OH, or H$_2$O~[4, 5]. Note that the absence of molecular and maser radiation is not rare 
for AGB stars (see, for instance,~[6]).

In our current study, we have aimed to determine the main parameters of this object and its evolutionary
status based on optical spectroscopy. We emphasize that, since no observations of the object in the optical
range are available, our spectral observations are the first for RAFGL\,5081. Section~2 brieﬂy describes
the observational data, and Section~3 presents information about the profiles of features detected
in high-resolution spectra. We discuss the results and attempt to apply them to determine luminosity,
distance, and evolutionary status of the star. The results are further discussed in Section~4, and the
main conclusions are summarized in Section~5.

\section{Observational data and reduction} 

A program of spectroscopic studies of AGB and post-AGB supergiants with IR excesses, including
some related stars with unclear evolutionary status, has been carried out on the 6-m telescope of the
Special Astrophysical Observatory (SAO) over the past two decades. The initial aim of the program was
to deﬁne the fundamental parameters of the studied objects and to search for anomalies in the chemical
compositions of their atmospheres associated with the synthesis of elements in the preceding evolutionary 
stages. The set of information obtained makes it possible to reliably establish the evolutionary status
of the star. During the realization of the program, the need arose for additional research aimed at 
identifying peculiarities in the spectra and variability of spectral features with time. A detailed 
study of the velocity field in the atmospheres and envelopes of the program stars was also necessary.

In the present study, we used spectra obtained at the Nasmyth focus using the NES spectrograph~[7],
which, in combination with a 2048$\times$2048 CCD array and image scanner, provides a spectral resolution
of R$\approx$60\,000. Since 2011, the NES spectrograph has been equipped with a large-format (2048$\times$4096)
CCD array, providing a significant increase in the simultaneously detected wavelength range. 
The details of the spectrophotometric and position measurements of the spectra are described in earlier papers,
which are referenced in~[2]. 

Table 1 presents information on the spectral data. To track the stability of these data over the 14--year
interval, we present heliocentric radial velocities Vr of several interstellar lines and bands: three components
of the Na\,I doublet and the narrowest DIBs at 5797, 6197, 6379, and 6614\,\AA{} (values obtained from the first 
band only are followed by colons). The two last rows of this table present the mean values of Vr and their
uncertainties. The close values of Vr in the last two columns conﬁrm the interstellar origin of the longest
wavelength component of the Na\,I doublet. The spectrophotometric and position measurements were
carried out using 24 spectra obtained on arbitrary dates in 2001--2015. The data for nearby dates were
averaged, and data are accordingly represented in Table~1 by 20 epochs, and in Table~2 by 21 epochs.

\section{Main features of the optical spectrum}

The optical spectrum of the counterpart of RAFGL\,5081 in the detected wavelength range
(3950--8480\,\AA{}  is an absorption spectrum. Most of the lines belong to ions and neutral 
atoms of iron-group elements; weak emission is present only in the long-wavelength wing of 
the H$\alpha$ line (see Fig.~1). Figure~2 compares a small fragment of the spectrum
with a corresponding spectral fragment for the cool supergiant HD\,246299, identified 
with the IR source IRAS\,05381+1012. The spectrum of HD\,246299 was obtained using the 
same NES spectrograph and processed using the same procedures. Its spectral type was determined 
to be G2\,I~[8]. This comparison of the spectra of the two stars shows that their spectral 
types are similar. The spectra also display similar line widths. Thus, in a first approximation, 
we can assume that the spectral type of the star is $approx$G5--8\,II.
This estimate of the spectral type is consistent with the estimated effective temperature of the 
star, Teff\,=5400$\pm$100\,K, which we determined using the set of criteria of Kovtyukh et al.~[9]. 
A late spectral type for the optical counterpart of RAFGL\,5081 is consistent with the absence 
of signatures of molecular and maser radiation from the stellar envelope. This indicates that
the star has not yet begun its evolution away from the AGB. We note here the study of Lewis~[10], who
examined the chronology of the appearance of various masers in post--AGB stars.

A high luminosity for the star is indicated by the blue-shifted wind components of the Na\,I\,(1), 
H$\alpha$, and H$\beta$ lines, whose profiles are presented in Fig.\,3, and by an absence of broad 
wings in the photospheric H$\beta$ and H$\gamma$ lines. A well-known criterion for stellar luminosity 
is the equivalent width of the O\,I\,7773\,\AA{}  triplet, whose calibration was obtained,
e.g., in~[11]. The equivalent width of this triplet in the spectrum of the counterpart of RAFGL\,5081 
is W(OI)\,=\,0.84\,\AA{}, implying the luminosity M$_{bol}$\,=$-4^m$. This estimate is not an accurate 
determination of the stellar luminosity, however, as the measured value of W(OI) should be considered 
a lower limit, since the individual components of the triplet overlap due to their large widths. 
In addition, the equivalent width of the triplet can be affected by the oxygen in the stellar 
atmosphere, whose abundance we do not know apriori. The moderate luminosity enables
us to reject a massive evolved star (supergiant or hypergiant); a low-mass supergiant in  the AGB stage 
is a more likely candidate for RAFGL\,5081. The large widths of the metallic absorption lines confirms  
also a high luminosity for the star. However, since the luminosity estimate based on the equivalent width of
the W(OI) triplet is not extremely high, the width of the metallic absorption lines is more likely related 
to macroturbulence and/or the axial rotation of the star.

The wind components of the Na\,I~D and H$\alpha$ profiles extend to short wavelengths, to velocities of
$-250$ and $-600$\,km/s, respectively, as is clearly show in Figs.\,1 and 3. Narrow components are clearly
visible in these profiles, whose number, depth, and positions vary with time. The limits of the time
variation of the Na I line can also be judged from Fig.\,4. The radial velocities for the cores of the 
components are presented in the third column of Table~2. In one of the spectra obtained on October 28, 2015,
which included red wavelengths, the parameters of two interstellar components of the K\,I~7696\,\AA{}
were measured. The locations and relative depths of both component are shown in Fig.\,4. It is obvious
that the expanding envelope of RAFGL\,5081 is inhomogeneous, since up to four clumps are present in
the line of sight simultaneously. This inhomogeneity and the asymmetry of the envelope are also manifest 
through appreciable instability of the positions of the wind components of the Na\,I~D and H$\alpha$ line 
profiles. These features of the Na\,I~D and H$\alpha$ line profiles suggests that the wind is non-stationary.

\begin{table}[ht!]
\caption{Spectral data and measured heliocentric radial velocities Vr for the interstellar Na\,I~D lines and DIBs. 
   The last two rows list the corresponding mean velocities Vr(aver) and their uncertainties.} 
\medskip                                                             
\begin{tabular}{ l|  l | l  l l l}
\hline
Date&  $\Delta\lambda$, \AA{} &\multicolumn{4}{c}{Vr, km/s}   \\   
\cline{3-6}
&        &\multicolumn{3}{c}{NaI} &\hspace{-0.1cm} DIBs \\   
\cline{3-6}
01.12.01 &  5030--6680  & --65.3  &--46.0 & --8.9 & --9   \\  
02.12.01 &  4560--6070  &         &       &       & --10: \\  
02.12.02 &  4520--5990  & --64.5  &--45.8 & --8.5 &      \\   
22.02.03 &  5150--6660  & --65.0  &--45.0 & --8.1 &--9   \\   
12.04.03 &  5270--6760  & --66.4  &--45.1 & --8.5 &--9   \\   
23.05.03 &  5270--6760  & --67:   &--45.0 & --8.0 &--10  \\   
17.08.03 &  5270--6760  & --64.5  &--45.5 & --9.0 &--10  \\   
11.09.03 &  5420--6000  & --63.7  &--44.2 & --8.0 &--6:  \\   
08.03.04 &  5280--6770  & --65.0  &--44.6 & --8.7 &--11  \\   
12.11.05 &  4560--6010  & --66.5  &--44.2 & --9.5 & --9  \\   
14.11.05 &  5280--6780  &         &       &       & --10 \\   
14.01.06 &  5280--6780  & --66.0  &--44.0 & --8.5 & --8  \\   
13.03.06 &  3980--5460  &         &       &       &      \\   
13.08.06 &  4560--6010  & --65.8  &--44.2 & --8.8 & --8  \\   
12.10.06 &  5220--6690  & --66.3  &--44.2 & --8.8 &--11  \\   
04.12.06 &  5220--6690  & --66.0  &--44.0 & --8.3 &--10  \\   
03.02.07 &  5220--6690  & --67.0  &--45.0 & --8.7 &--10  \\   
07.02.07 &  4560--6010  &         &       &       & --6: \\   
08.03.07 &  4560--6010  & --67.5  &--45.5 & --8.5 &--7   \\   
16.01.08 &  5210--6680  & --65.8  &--44.5 & --8.5 &--10  \\   
04.11.08 &  4460--5930  & --65.9  &--43.6 & --8.8 &--6:  \\   
18.04.14 &  3950--6980  & --62.0  &--42.0 & --8.0 &--9   \\   
30.09.14 &  5400--8480  & --62.0  &--43.0 & --8.6 &--9   \\   
28.10.15 &  3950--6980  & --63.0  &--43.0 & --9.0 &--11  \\   
 \cline{3-6}
         & V(aver)=    &$-65.3$  &$-44.4$&$-8.5$ &$-9.3$ \\    
&&\hspace{0.1cm} $\pm0.4$ &\hspace{0.1cm} $\pm0.2$&\hspace{0.1cm}  $\pm0.1$&$\pm0.3$ \\
\hline
\end{tabular}                                   
\end{table}

\begin{table}[ht!]
\caption{Heliocentric radial velocities Vr of the absorption components of the line profiles. The Vr values in the 2-nd
        column were derived from lines of iron-group metals.}
\medskip                                                             
\begin{tabular}{l| c c l| r r  r| l r| c c r l}
\hline
             & \multicolumn{12}{c}{V${\rm _r}$, km/s } \\
\hline             
       Дата  &\multicolumn{3}{c|}{Metals}&\multicolumn{3}{c|}{Na\,I}  &\multicolumn{2}{c|}{H$\beta$} &\multicolumn{4}{c}{H$\alpha$} \\
\hline
    1,2.12.01 & $-52$& $(-21)$&  9   &$-139$ &$-88$ &      &  $-160$&       &   $-350$& $-225$& $-164$ &       \\   
     02.12.02 & $-54$& $(-22)$&  8   &$-120$ &$-80$ &      &  $-200$&$-120$ &         &       &        &       \\   
     22.02.03 & $-56$& $(-21)$& 11   &$-110$ &$-78$ &      &        &       &   $-360$& $-130$&        &       \\   
     12.04.03 & $-54$& $(-21)$& 11   &$-132$ &$-82$ &      &        &       &   $-260$& $-115$&        &        \\  
     23.05.03 & $-51$& $(-22)$& 10   &$-160$ &$-10$3&$-78$ &        &       &   $-400$& $-335$& $-180$ &$-87$   \\  
     17.08.03 & $-55$& $(-23)$& 11   &$-105$ &$-81$ &      &        &       &   $-200$& $-120$&        &        \\  
     11.09.03 & $-50$& $(-20)$& 11   &$-102$ &$-81$ &      &  $-128$&       &         &       &        &        \\  
     08.03.04 & $-53$& $(-22)$& 10   &$-135$ &$-11$3&$-76$ &        &       &   $-320$& $-210$& $-148$ &$-85 $  \\  
12,14.11.05   & $-53$& $(-22)$& 10   &$-135$ &$-83$ &      &  $-135$&       &   $-485$& $-375$& $-245$ &$-146$  \\  
     14.01.06 & $-52$& $(-21)$& 10   &$-135$ &$-80$ &      &        &       &   $-280$& $-140$& $-90$  &         \\ 
     13.03.06 & $-53$& $(-21)$&  6   &       &      &      &  $-150$&       &         &       &        &         \\ 
     13.08.06 & $-52$& $(-21)$& 10   &$-148$ &$-11$4&$-105$&  $-150$&       &         &       &        &         \\ 
     12.10.06 & $-53$& $(-21)$& 11   &$-172$ &$-10$8&      &        &       &   $-290$& $-188$& $-123$ &         \\ 
     04.12.06 & $-54$& $(-21)$& 13   &$-164$ &$-12$1&$-96$ &        &       &   $-260$& $-166$& $-126$ &$-100$   \\ 
    3,7.02.07 & $-55$& $(-22)$& 11   &$-190$ &$-14$6&$-100$&        &       &   $-195$& $-100$&        &         \\ 
     08.03.07 & $-52$& $(-22)$& 11   &$-130$ &$-95$ &      &  $-128$&       &         &       &        &         \\ 
     16.01.08 & $-54$& $(-21)$& 11   &$-173$ &$-13$2&$-100$&        &       &   $-180$& $    $&        &         \\ 
     04.11.08 & $-55$& $(-22)$& 10   &$-174$ &$-11$6&$-93$ &  $-170$&$-115$ &         &       &        &         \\ 
     18.04.14 & $-51$& $(-22)$&  5   &$-188$ &$-13$0&      &  $-193$&       &   $-280$& $-200$&        &         \\ 
     30.09.14 & $-52$& $(-21)$&  9   &$-124$ &$-87$ &      &        &       &   $-300$& $-135$&        &         \\ 
     28.10.15 & $-51$& $(-21)$&  9   &$-134$ &$-10$4&      &  $-180$&       &   $-200$& $-145$&        &         \\ 
\hline
\end{tabular} 
\end{table}

The profiles of photospheric metallic absorptions are also unusual. Nearly all of these are split,
and, unlike the wind absorption lines, are stationary. The lines in the spectrum of RAFGL\,5081 are broad
and strongly blended, but the splitting of weak absorptions makes their components distinguishable 
(e.g., the blue components of Ti\,I\,6126 and Fe\,I\,6128\,\AA{} in Fig.\,2). Other examples are shown
in Fig.\,5, which presents the profiles of several lines of varying strength, from the weak Fe\,I~6042\,\AA{} 
line to the TiII+FeII~4549\,\AA{} line. This clearly shows that blue and red components approach each 
other as the lines become deeper, until the profile becomes that of a single line.

The variation of the absorption profiles with their depth, shown through the dependence of the radial
velocity on the residual intensity, is illustrated in Fig.\,6. The data here were averaged over the spectra
obtained on April~18, 2014 and October~28, 2015. Corresponding Vr(r) dependences were constructed
for each spectrum and used to estimate the radial velocities for both absorption components (on average,
at about $-54$ and +10\,km/s). These are presented in the second column of Table~2 on either side of the
parantheses containing the central velocity values, which were derived from the bisector of the profile as
a whole. Figure~6 and Table~2 show that the central velocities do not change with either line depth or with
time.

The constance of the split lines excludes the possibility that the splitting is due to binarity. This is all the
more so as the binary components would have to be stars with close values of their absorption intensities.
It is also not possible to obtain the observed profile shape by superposing absorption and emission lines:
on average, the depths of the lines in the spectrum of RAFGL\,5081 are the same as those of HD\,246299,
and their equivalent widths are even larger.

Estimation of the velocity of the center of mass of the star (Vsys) is related to estimation of its distance.
As we noted in the Introduction, neither molecular nor maser radiation from RAFGL 5081 are detected.
This makes it natural to suppose that Vsys is close to the average velocity of weak absorption features,
Vsys$\approx -21$\,km/s. Velocity estimates based on the photospheric components of the H$\beta$, H$\gamma$, 
and H$\delta$ profiles and the H$\alpha$ emission component also support this hypothesis (see Fig.\,1). 
It would also be useful to take into account C$_2$ and CN molecular absorption, which can form in the 
outer envelopes of cool stars, but we could not identify such features in our spectra.
It may be that these molecular absorption lines will be found in spectra with higher resolution and higher
signal-to-noise ratio.

Let consider the narrow absorption features in the Na\,I\,(1) doublet in more detail. It seems that only the
two components of Na\,I\,(1) with velocities of $-8.5$ and $-65.3$\,km/s, whose positions coincide with the
the position of the components of the K\,I\,7696\,\AA{}  line, have an interstellar origin. Within the 
uncertainties, the average velocity estimated from diffuse interstellar bands, Vr(DIBs)\,=$-9.3$\,km/s, agrees with
the position of the interstellar Na\,I\,(1) component with Vr(IS)\,=$-8.5$\,km/s, which forms in the local
Galactic arm. Taking into account the data of~[12] for the Galactic structure, the presence of components 
with the velocity $-65.3$\,km/s in the Na\,I\,(1)  profiles suggests that the star, which has Galactic 
coordinates l/b$\approx 133\degr/9\degr$, is located behind the Perseus arm and is therefore no closer 
than 2\,kpc. 

Adopting for RAFGL\,5081 the above estimate Vsys$\approx -21$\,km/s, the velocity of the star appears peculiar. 
It cannot be ruled out that the third component with the average velocity Vr\,=$-44.4$\,km/s forms in 
the circumstellar medium. In this case, we can estimate the expansion velocity of the envelope as 
Vexp$\approx 23$\,km/s. Comparing this value of Vexp with the data for the sample of post-AGB stars 
studied in~[2] shows that this derived velocity is typical for stars in the AGB and post--AGB stages.

\section{Discussion of the results} 

The limited nature of the observational data available for RAFGL\,5081 hinders reliable determination
of its fundamental parameters, accurate calculation of its chemical composition, and thus determination of
its evolutionary status. One natural way to at least partially solve these problems is to consider searches
for similar objects among stars studied earlier. The long-term program of spectroscopy of anomalous 
supergiants with large IR excesses carried out on the SAO 6-m telescope resulted in an extensive 
collection of high-quality spectra of objects of this type. We were able to draw some  
conclusions concerning evolutionary changes in their fundamental parameters, their detailed 
chemical compositions, and peculiarities of the kinematic states of their atmospheres and structured 
extended circumstellar envelopes.

In particular, the analysis of the extensive collection of high-quality echelle spectra of post--AGB stars
enabled the detection of previously unknown features in the spectra of selected post-AGB stars with 
atmospheres enriched in carbon and heavy s-process metals and circumstellar envelopes containing 
carbon molecules: splitting (or asymmetric) strong absorption lines~[2, 13]. Using published IR 
and radio spectroscopic data, it was shown that the individual components of the split absorptions  
are formed in the structured circumstellar envelopes of post-AGB stars~[2]. In this way, observational 
evidence for enrichment of the circumstellar envelopes in heavy s-process metals was obtained. 
The splitting into components of the profiles of the strongest heavy-metal absorption features 
found in the spectra of the post--AGB supergiants V5112\,Sgr~[14], V354\,Lac~[15],
and CGCS 6918 [16], which have extended envelopes with complex morphologies, suggests that the 
formation of a powerful structured envelope in the AGB stage is accompanied by an outﬂow of stellar 
nucleosynthesis products into the interstellar medium.  Attempts to find a direct relationship between 
features of the optical spectrum and the morphology of the circumstellar medium are complicated by the fact
that the observed structure of the envelope depends strongly on the inclination of the symmetry axis to
the line of sight and the angular resolution of both the spectral and imaging instrumentation.

Earlier, in our program of spectroscopy of supergiants with IR excesses, we studied the F star V2324\,Cyg, 
which displays some spectral features that place it outside the sample of canonical post--AGB stars. 
The most important of these is the presence of strong and variable H$\alpha$ emission~[17] in the spectrum 
of V2324\,Cyg, which contains a wind component with a velocity of up to 450\,km/s. Other lines of neutral 
hydrogen and the Na\,I~D lines also have complex P~Cygni-type profiles indicating high wind
velocities. Moreover, the profiles of metal absorption lines in the spectrum of V2324\,Cyg are broadened
by the high microturbulence and rotational velocities: $\xi_t$\,=\,6\,km/s and Vrot\,=\,69\,km/s, 
respectively~[17].

V2324\,Cyg is probably a massive star in the post-AGB stage, with an initial mass of M$_{\rm init} \ge $4\,M$_{\sun}$,
judging from the very large excesses of Li and Na in its atmosphere. Excess abundances of light metals
and lithium can be the result of effective hot bottom burning in the course of the evolution of the most
massive AGB stars (see, e.g., [18]). It is important that fairly strong Li\,I~6707\,\AA{}  absorption has 
also been reliably identified in spectra of the optical counterpart of RAFGL\,5081, with an equivalent 
width averaged over five spectra of W$_{\lambda}$\,=\,106\,m\AA{}. The presence of this line indicates 
that RAFGL\,5081 is among the more massive AGB stars, in good agreement with its association with OH/IR 
sources~[1]. Using our estimate of Teff, a surface gravity typical for cool supergiants, $\log g$\,=\,1.0, 
and the microturbulence velocity  $\xi_t$\,=\,6\,km/s, we obtained for   W$_{\lambda}$\,=\,106\,m\AA{}
the fairly high lithium abundance $\log\epsilon{\rm (LiI)}\approx2.6$, which should be considered a 
preliminary estimate only.

The optical component of the OH/IR star IRAS\,18123+0511 was studied for the first time using spectra 
obtained with the SAO 6-m telescope~[19]. Like in the case of RAFGL\,5081, IRAS\,18123+0511 was not 
identified with any known optical object, and no photometric measurements are available. According 
to~[19], the abundance of iron-group elements (V, Cr, Fe) in the atmosphere of the central star of 
IRAS\,18123+0511 is one-third the solar value. The calculated abundance ratio O/C$>$1 is consistent with 
the existence of silicate emission at a wavelength of about 9.7\,$\mu$ in the IR spectrum of this object. 
The presence of such a feature is characteristic of AGB stars with oxygen-enriched atmospheres. 
The lithium 6707\,\AA{} doublet was measured quite confidently in the spectra of IRAS\,18123+0511, 
but the calculated lithium abundance is extremely low. 
This can be explained by the destruction of lithium under the conditions of convection. 
The abundances of s-process heavy metals is low compared to the metallicity: their average
values are 0.6 the solar value. In general, the pattern of atmospheric abundances in IRAS\,18123+0511 is
consistent with our concept of a massive star in the early AGB stage. Its metallicity, combined with its
radial velocity and Galactic latitude, indicates that it belongs to the old Galactic-disk population.

It is interesting to compare our results for RAFGL\,5081 with those obtained earlier for the cool
supergiant V1027\,Cyg, which is the optical counterpart of IRAS\,20004+2955. That source also lacks
radio spectroscopy, however, SAO 6-m spectra were used to determine the fundamental parameters of the
atmosphere of the central star and the abundances of a large number of elements~[20]. Later, based
on the results of monitoring V1027\,Cyg with the NES spectrometer on the 6-m telescope, a feature
that was previously unknown for stars close to the AGB was detected in its spectrum~[21]. Klochkova
et al.~[21] discovered splitting of the cores of strong absorption lines of metals and their ions 
(Si II, Ni I, Ti I, Ti II, Sc II, Cr I, Fe I, Fe II, Ba II). It turned out that the broad profiles 
of strong lines contain a stable weak emission feature in the line cores, which splits the profile 
into components. It was shown that the mean position of the emission features can be
considered to be a systemic velocity, Vsys\,=\,5.5 km/s.
In addition, symmetric weak and moderate absorption features displayed slight changes in radial velocity
with an amplitude of 5--6\,km/s, due to pulsations. The results of radial velocity measurements based
on the positions of the components of the strongest absorption lines in the spectrum of V1027\,Cyg were
a mystery. The accuracy of measurements of the position of the full profile of a split absorption line
is lower than the accuracy of the components of the profile itself due to asymmetry, but this accuracy is
sufficient to draw conclusions about the behavior of the position of the profile with time. The position of
the complete profile remains essentially constant, and does not correspond to the velocity derived 
from weak absorption features. This can be understood if strong lines forming in the outer layers of 
the atmosphere are stationary, and are not affected by pulsations. These considerations lead to the 
conclusion that the circumstellar envelope of V1027\,Cyg is fairly stable.

Studying the spectra of the optical counterpart of  RAFGL\,5081, we again see the splitting of absorption features. However, 
as follows from the discussion in the previous sections, the properties of the optical spectrum of RAFGL\,5081 differ 
from those detected in other stars with IR excesses. Currently, with the very limited observational data available for 
this object, we cannot identify firm explanations for its spectral features, but we can make some guesses. Judging 
from the IR ﬂuxes measured in~[1], RAFGL\,5081 is located in region IV on the [12-25]--[25-60]  IR-color diagram 
presented in~[22], which contains stars with high mass-loss rates and optically thick circumstellar envelopes. 
In addition, the spectral energy distribution of RAFGL\,5081~[1] indicates that the envelope has not yet separated 
from the central star. Since the temperature of the central star is low, and is not sufficient to excite the emission
of the nebula, the nebula radiates due to scattering of the radiation of the star. This suggests that we are observing 
an extended and outﬂowing pseudo-photosphere.

As in the case of the V1027\,Cyg spectrum, it would be natural to suggest envelope emission formed near the surface 
layers of this pseudo-photosphere as a factor leading to the splitting of the absorptions in the spectrum of RAFGL\,5081.
The position of this emission would coincide with the velocity of the emission component in the long-wavelength wing of 
the H$\alpha$ line, which can be clearly distinguished at some times in Fig.\,1. The large
widths of the emission and absorption lines can be explained by both rotation and a high macroturbulence velocity in the 
pseudo-photosphere. However, the presence of envelope emission features is not consistent with the fact that the emission 
intensity in the RAFGL\,5081 spectrum is independent of the excitation potential of the line.  One way to explain 
the splitting of the absorption lines is to suppose the formation of a large-scale  structure that can be considered 
a predecessor of  future rotating torus, disk, or pair of lobes in the extended pseudo-photosphere of the cool supergiant.
Freytag et al.~[23] performed simulations of the complex structure of the surface observed with a high  spatial resolution 
for nearby AGB stars. In particular, they noted that the origin of the asymmetry of images of AGB stars could be deep 
convection in the form of giant, long-lived convective cells.

In essence, the observed structure of a proto-planetary nebula shell is a record of the evolution of the central star’s 
mass loss. There currently exist successful theoretical models that can explain the formation of such a complex morphology 
during the evolution of a single AGB star. K\"oning et al.~[24] suggested a PPN model based on a pair of cavities of 
reduced density inside a dense spherical halo, and showed that it is possible to reproduce all the morphological features 
observed in real bipolar PPNs through variation of the model parameters (density, dimensions, and orientation of the cavity)

The need for continued studies of the IR source RAFGL\,5081 using high-resolution spectroscopy on the largest telescopes, 
in order to determine the parameters of the atmosphere model and calculate the elemental abundances, is clear to us. 
Observations with high spatial resolution, as well as photometric monitoring, radio spectroscopy, and spectropolarimetry, 
would also be useful. Spectropolarimetry with high spectral resolution would enable clarification of the positions 
of areas of formation of different spectral features, and, consequently, of the structure of the RAFGL\,5081 system. 
The results of spectropolarimetric observations of the IR source RAFGL\,2688 [25] are relevant here. The circumstellar 
envelope of that object has a complex structure due to bipolar and jet outﬂows from the central post-AGB star, 
with its outer, spherical envelope formed in the previous AGB stage. The complex structure of such an envelope begins 
to form after the transition from a spherical wind in the AGB stage to a superwind in the subsequent stage~[26]. 
To develop possibilities for polarimetric observations with high spectral resolution,  an echelle spectropolarimeter 
designed to operate at the primary focus of the SAO 6-m telescope is currently in the assembly stage~[27, 28].

Our interpretation of the multicomponent Na\,I\,(1)  profiles is supported by the presence of Na\,I\,(1) components 
with the same velocities in the spectra of several distant Galactic stars located in similar directions. For example, 
as part of the PPN spectroscopy program, we studied the optical spectrum of IRAS\,23304+6147 in detail~[29] (Galactic 
coordinates l/b$\approx 113\degr/0\degr$). This remote object belongs to a small subgroup of PPN with atmospheres 
enriched in s-process products~[13]. In the current context, it is important that an interstellar  component of 
the Na\,I\,(1) lines with the velocity Vr(IS)\,=$-61.6$\,km/s was identified in the optical spectrum of  
IRAS\,23304+6147~[16], similar to the case of RAFGL\,5081. Moreover, the spectrum of this object also contains 
a circumstellar component of the Na\,I\,(1) lines with velocity Vr\,=$-41.0$\,km/s, close to the velocity observed 
in the spectrum of RAFGL\,5081. The circumstellar origin for this component in the spectrum of IRAS\,23304+6147 is 
confirmed by the excellent coincidence with the velocity derived from the rotational lines of the C$_2$ Swan bands. 
A similar pattern of interstellar components is also detected in the spectrum of VES\,695  -- the optical counterpart 
of IRAS\,00470+6429~[30], with Galactic coordinates l/b$\approx 122\degr/2\degr$. As was shown by Miroshnichenko 
et al.~[30], this B[e]-type star is located at a distance of $\approx$2\,kpc, and is probably a member of the 
Cas\,OB7 association.

Consideration of the small number of available publications concerning RAFGL\,5081 suggests that its optical 
counterpart is an unnamed weak star mentioned in the catalog of van den Berg~[31], where it is noted that 
reflection nebula No. 7 is associated with a weak star to the south of RZ\,Cas. This nebula is designated 
GN\,02.44.7 in the modern catalog of reflection nebulae~[32]. According to SIMBAD, the coordinates of 
RAFGL\,5081 and nebula GN\,02.44.7 coincide.

\section{Conclusion}

Based on long-term (2001--2015) spectral monitoring with high resolution performed on the SAO 6-m telescope 
using the NES spectrograph, we have obtained for the first time optical spectra of the optically weak IR source 
RAFGL\,5081.

Based on a comparison with related objects, the spectral type of the star is close to G5—8\,II. This spectral 
type is consistent with the effective temperature of the star, Teff\,=\,5400$\pm$100\,K, estimated using 
spectral criteria. The equivalent width of the O\,I~7773\,\AA{}  oxygen triplet in the spectrum of the optical 
counterpart, W(OI)\,=\,0.84\,\AA{}, provides the preliminary estimate of the luminosity of RAFGL\,5081 M$_{bol}=-4^m$. 
This low luminosity rules out the possibility that the object is a massive red supergiant. Li\,I~6707\,\AA{}  
absorption is reliably identified in the spectrum of the counterpart of RAFGL\,5081; its equivalent width,
W$_\lambda\approx 0.1$\,\AA{}, testifies that this object is a relatively massive AGB star with initial mass 
M$_{\rm init}\ge 4$M$_{\sun}$.

The heliocentric radial velocities Vr corresponding to the positions of all components of metallic absorption lines, 
as well as the Na\,I~D and H$\alpha$ lines, have been measured for all the observation dates. Analysis of the 
multicomponent profiles of the Na\,I\,(1) lines revealed the presence of components with Vr(IS)\,=$-65$\,km/s. 
Based on this, we concluded that RAFGL 5081 is located in the Perseus  arm, i.e., no closer than 2\,kpc.

An unusual spectral phenomenon was found in the spectra of the star: the broad profiles of medium- and low-intensity 
absorption lines are split and, unlike the wind absorption lines, stationary. The constance ofthese absorption lines 
rules out that the anomalous absorption profiles are due to binarity.

The radial velocities of the wind components of the Na\,I\,D and H$\alpha$ lines reach values of $-250$ and $-600$\,km/s, 
respectively. These profiles contain narrow components, whose number, depths, and  positions vary with time. The time 
variability of the multicomponent structure of the Na\,I\,D and H$\alpha$ profiles indicate inhomogeneity and instability 
of the circumstellar envelope of RAFGL\,5081.

\section*{Acknowledgements} 
This study was supported by the Russian Foundation for Basic Research (project 16-02-00587\,a). 
This work has made use of the SIMBAD and ADS databases. Observations on the 6-m telescope of the Special Astrophysical 
Observatory are financially supported by the Ministry of Education and Science of the Russian Federation 
(contract 14.619.21.0004, project identifier RFMEFI61914X0004).

\begin{figure}[ht!]
\includegraphics[angle=0,width=70mm,bb=30 50 550 790,clip]{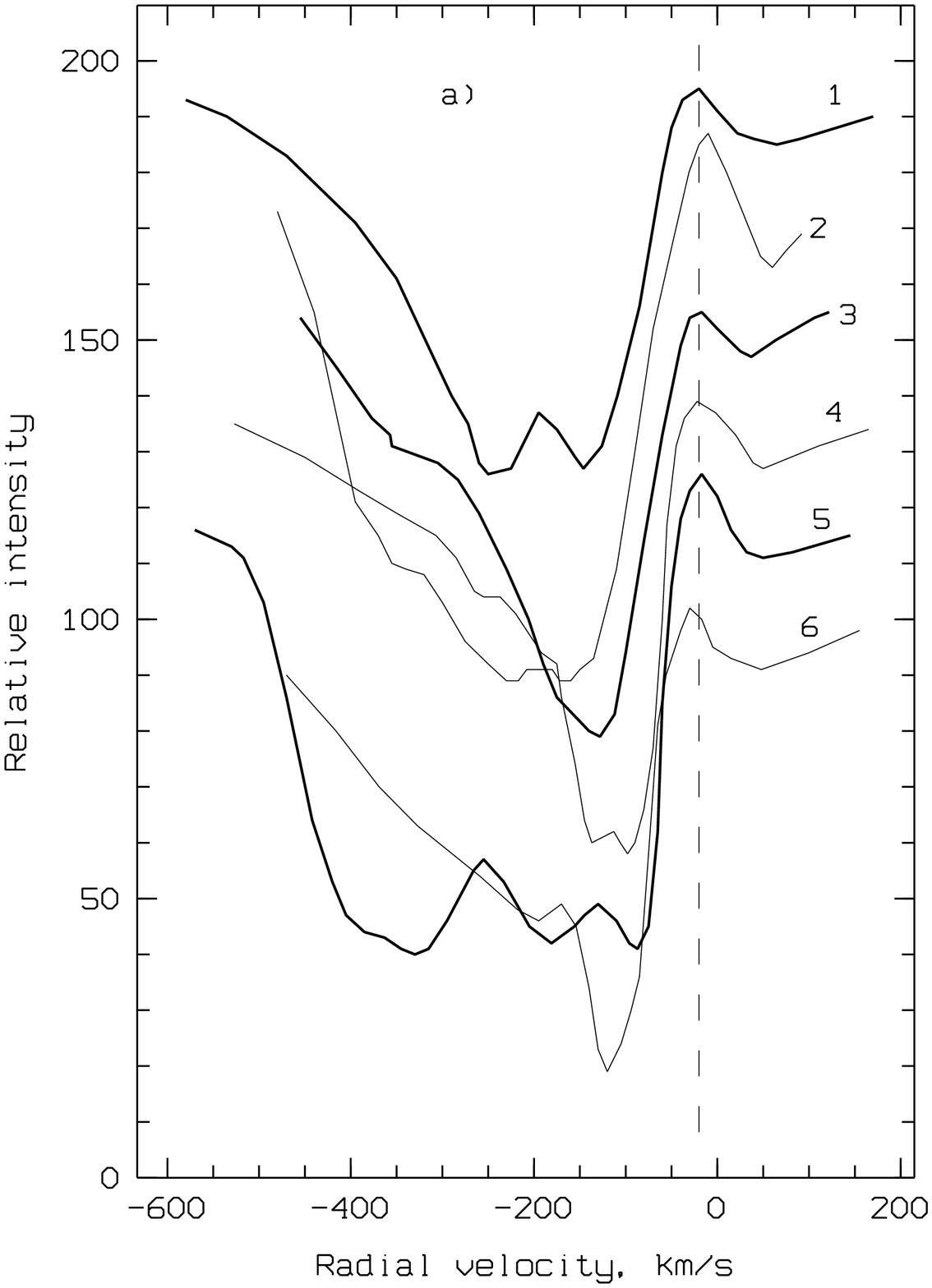}  %Halpha1.ps
\includegraphics[angle=0,width=70mm,bb=30 50 550 790,clip]{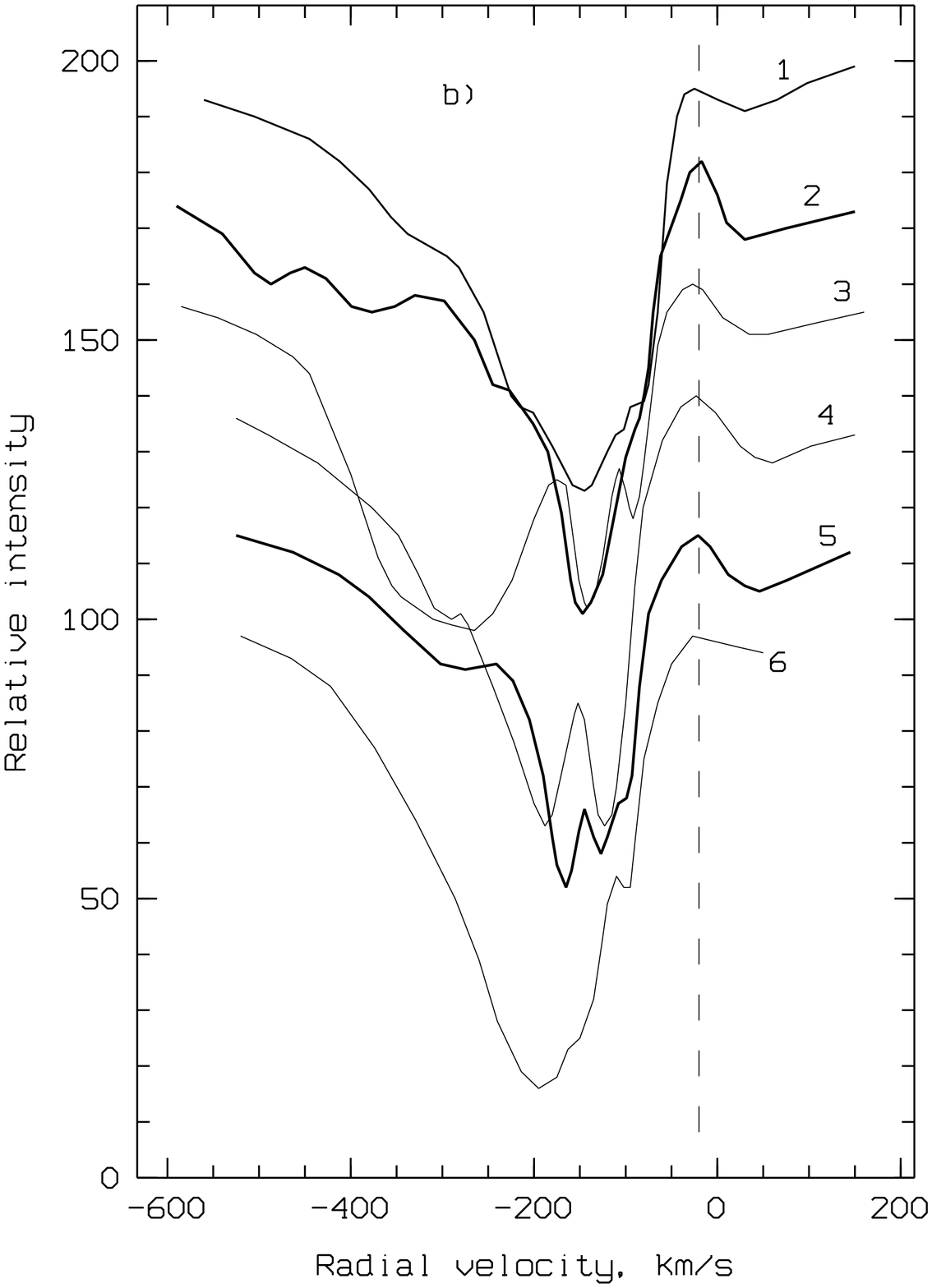}  %Halpha2.ps
\includegraphics[angle=0,width=70mm,bb=30 50 550 790,clip]{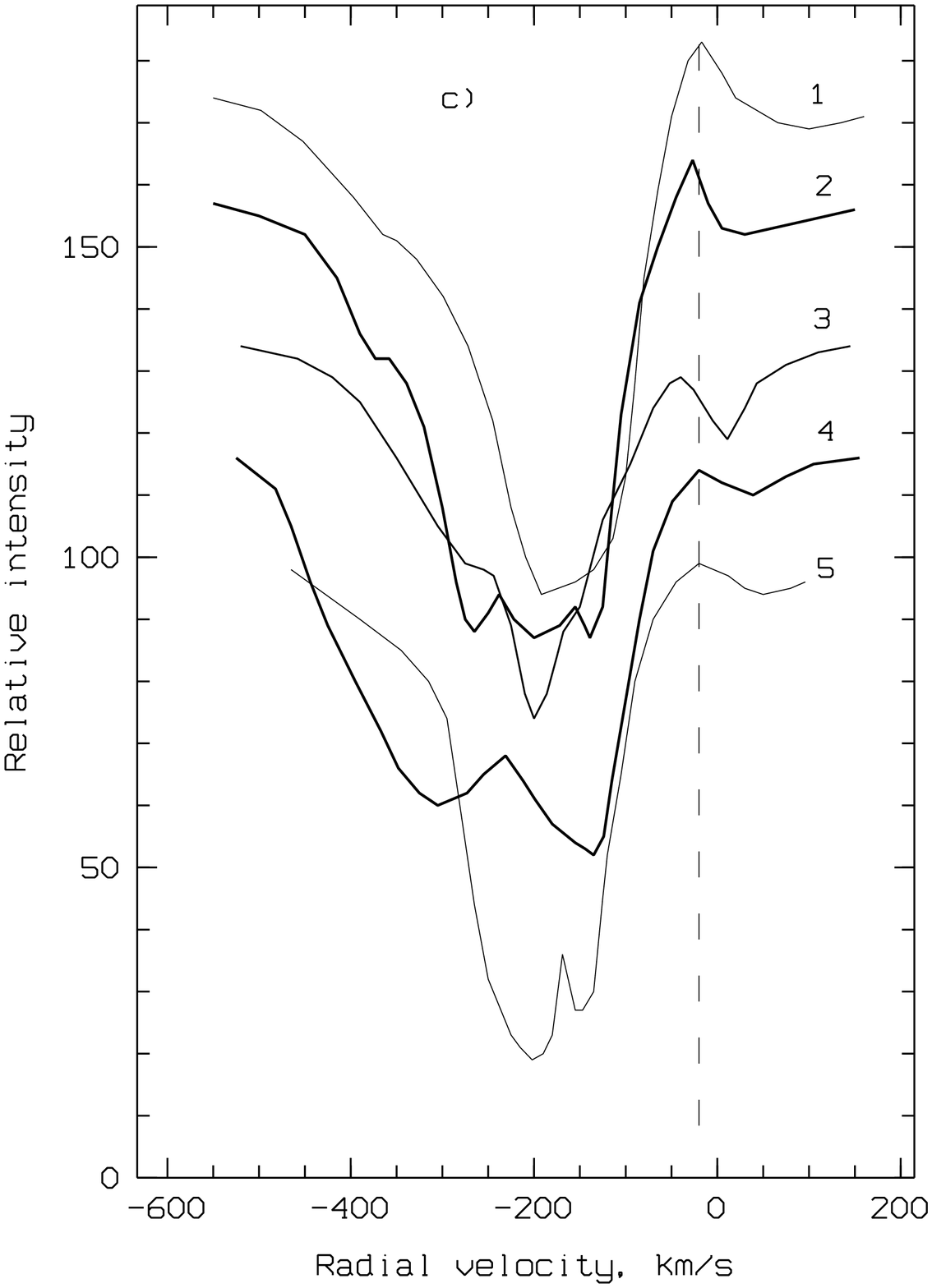}  %Halpha3.ps
\caption{H$\alpha$ profile on the following dates: (a) 1--March 7, 1999, 2--December 1, 2001, 3--February 22, 2003, 
   4--April 12, 2003, 5--May 23, 2003; (b) 1--August 17, 2003, 2--March 08, 2004, 3--November 11, 2005, 4--January 14, 2006, 
   5--October 12, 2006, 6--December 4, 2006; (c) 1--January 16, 2008, 2--October 11, 2013, 3--April 18, 2014,
   4--September 30, 2014, 5--October 28, 2015. The vertical dashed line shows the adopted systemic velocity Vsys=$-21$\,km/s
      (see the text). }
\label{Halpha}
\end{figure}

\begin{figure}
\includegraphics[angle=-90,width=110mm,bb=30 60 545 790,clip]{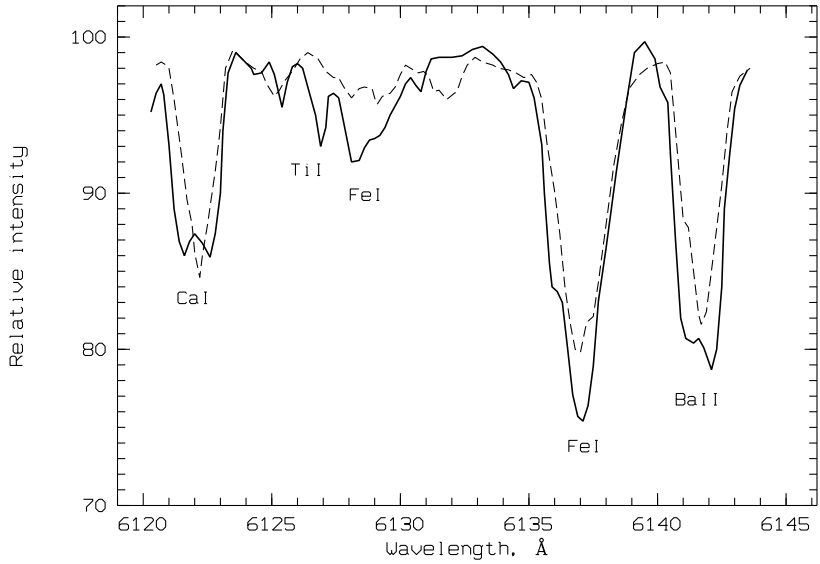} 
\caption{Comparison of fragments of spectra of two supergiants close to the AGB stage: RAFGL\,5081 (solid) and HD\,246299
         (dashed). Indicated wavelengths are their laboratory values. The labels show identifications of the main absorptions.}
\label{2Sp}
\end{figure}

%Fig. 3. 
\begin{figure}
\includegraphics[angle=-90,width=110mm,bb=20 60 550 790,clip]{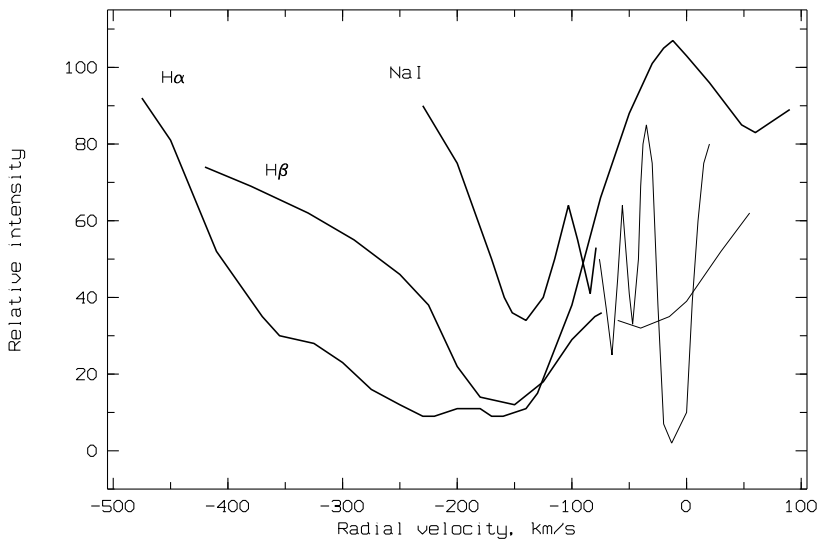} 
\caption{Profiles of the Na\,I~D2, H$\beta$, and H$\alpha$ lines in the spectrum of RAFGL\,5081 averaged 
        over two nearby dates: December~1 and December~2, 2001. The bold curves show wind components and thin curves 
        the interstellar part of the Na\,I profile and photospheric part of the H$\beta$  profile.}         
\label{Wind}
\end{figure}

%Fig. 4. 
\begin{figure}
\includegraphics[angle=-90,width=120mm,bb=30 80 550 790,clip]{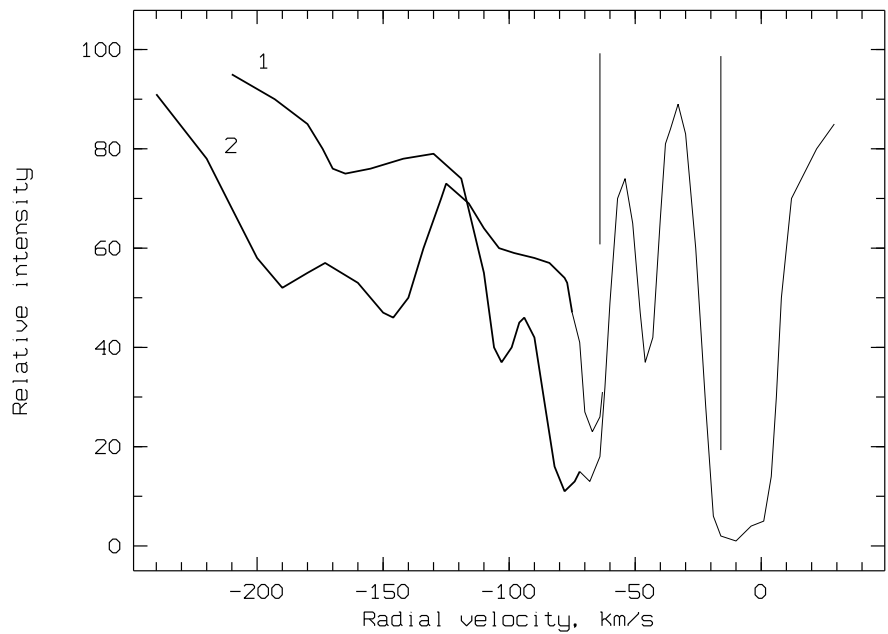} 
\caption{Time variations of the wind component of the Na\,I~D2 line profile (bold curve). Shown are the profiles 
        1--for  May 23, 2003 and 2--the mean profile February 3 and February 7, 2007. The interstellar and circumstellar 
        components are shown by the thin curves. The vertical solid bars show the positions of the two components of 
        the K\,I\,7696\,\AA{} line. The ratio of the heights of the vertical bars corresponds to the depths of these components.}         
\label{Na-var}
\end{figure}

\begin{figure}
\includegraphics[angle=0,width=70mm,bb=30 75 550 790,clip]{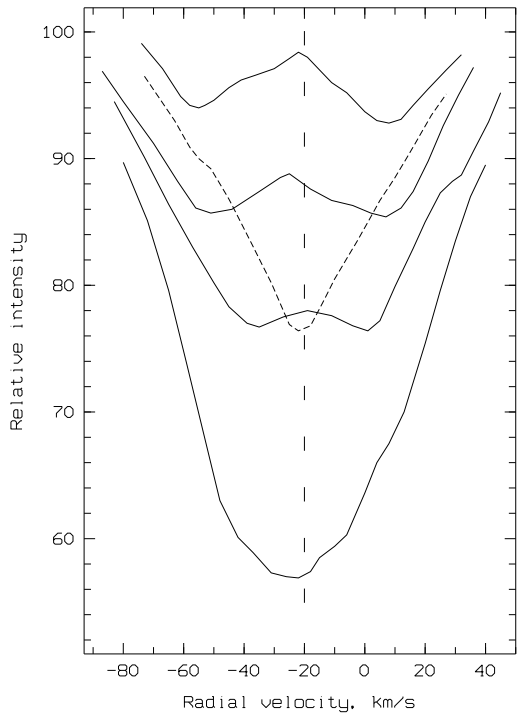}  % Profiles.ps
\caption{Examples of the profiles of photospheric absorptions in the spectrum of RAFGL\,5081, 
      shown by the solid curves (top to bottom): Fe\,I\,6042\,\AA{}, Ca\,I\,6122\,\AA{},  Fe\,II\,5317\,\AA{}, 
      and a blend Ti\,II+Fe\,II\,4549\,\AA{}. The short-dash curve shows the Fe\,II\,5317\,\AA{} profile in 
      the spectrum of HD\,246299. The vertical dashed line shows the adopted systemic velocity, Vsys\,=$-21$\,km/s.}
\label{Metals}
\end{figure}

\begin{figure}
\includegraphics[angle=-90,width=140mm,bb=30 50 550 790,clip]{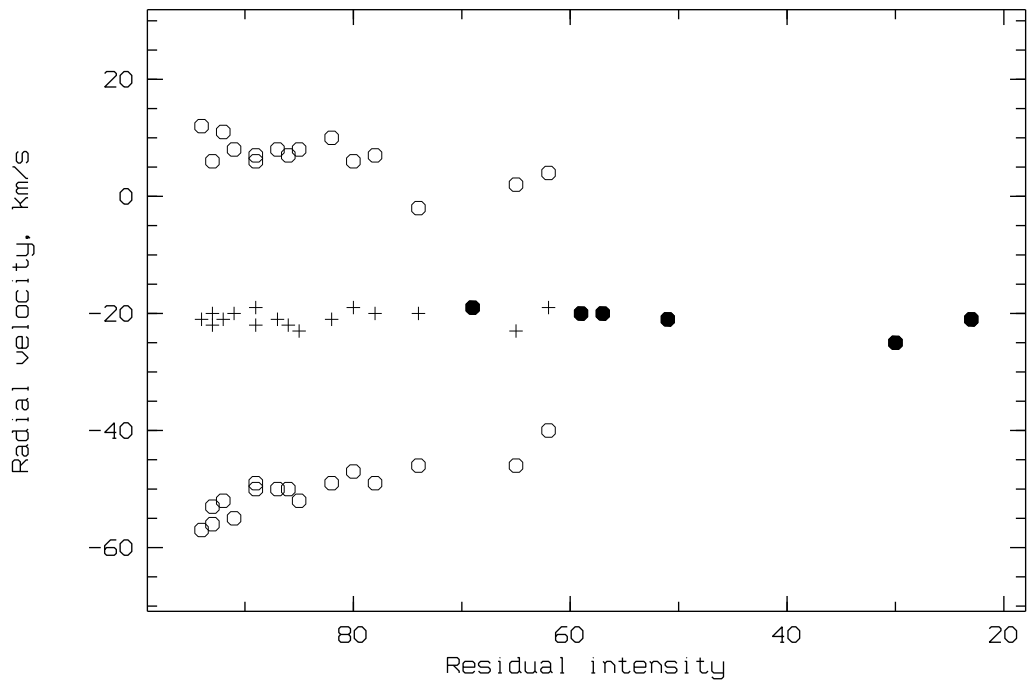} 
\caption{Dependence of the heliocentric radial velocities of photospheric absorptions on the residual 
        central intensity. The measurements were averaged over the spectra obtained on April~18, 2014 and 
        October~28, 2015. Every symbol corresponds to one line; open circles correspond to the blue and 
        red components, crosses to the profile centers, and filled circles to unsplit lines (the two 
        symbols at the right side represent the H$\delta$ and H$\gamma$ lines).}
\label{Vr-relation}
\end{figure}

\end{document}